\documentclass[aps,pra,twocolumn,groupedaddress,showpacs]{revtex4}
\usepackage{graphicx}
\usepackage{color}
\usepackage{amsmath}
\usepackage{amssymb}
\usepackage{txfonts}

\begin{document}
\title{Interference pattern and visibility of a Mott insulator}
\author{Fabrice Gerbier, Artur Widera, Simon F{\"o}lling, Olaf Mandel, Tatjana Gericke and Immanuel Bloch}
\affiliation{Institut f{\"u}r Physik, Johannes
Gutenberg-Universit{\"a}t, 55099 Mainz, Germany.}
\date{\today}
\begin{abstract}
We analyze theoretically the experiment reported in [F. Gerbier et
al, cond-mat/0503452], where the interference pattern produced by
an expanding atomic cloud in the Mott insulator regime was
observed. This interference pattern, indicative of short-range coherence in the system,
could be traced back to the presence of a small amount of
particle/hole pairs in the insulating phase for finite lattice
depths. In this paper, we analyze the influence of these pairs on the
interference pattern using a random phase approximation, and
derive the corresponding visibility. We also account for the
inhomogeneity inherent to atom traps in a local density
approximation. The calculations reproduce the experimental
observations, except for very large lattice depths. The deviation
from the measurement in this range is attributed to the increasing
importance of non-adiabatic effects.
\end{abstract}
 \maketitle
%
%
%
The superfluid to Mott insulator (MI) transition undergone by an
ultracold Bose gas in an optical lattice has attracted much
attention in the recent years as a prototype for strongly
correlated quantum phases
\cite{jaksch1998a,greiner2002a,zwerger2003a,stoferle2004a}. A key
observable in these systems is the interference pattern observed
after releasing the gas from the lattice and letting it expand for
a certain time of flight. Monitoring the evolution of this
interference pattern not only reveals the superfluid-to-MI
transition \cite{greiner2002a,stoferle2004a}, but also allows for
example the detection of number-squeezed states in the lattice
\cite{orzel2001a,hadzibabic2004a}, or the observation of collapse
and revivals of coherence due to atomic interactions
\cite{greiner2002b}. Because of its experimental importance, a
quantitative understanding of this interference signal is crucial
to characterize quantum phases of bosons in optical lattices.

Although no interference pattern is expected for a uniform array
of Fock states (what we call a ``perfect'' Mott Insulator)
\footnote{A recent experiment has shown that this is not
necessarily true in single-shot experiments for a large number of
atoms per site \cite{hadzibabic2004a}. Our experimental parameters
are such that no residual interference is expected, in agreement
with our observations.}, a finite visibility is nevertheless
observed in experiments above the insulator transition
\cite{greiner2002a,stoferle2004a,gerbier2005a}, in agreement with
numerical calculations \cite{kashurnikov2002a,roth2003a}. We have
studied this phenomenon experimentally, and shown that despite its
insulating nature that forbids long-range coherence, a MI still
exhibits short-range coherence at the scale of a few lattice sites
\cite{gerbier2005a}. This can be attributed to the structure of
the ground state for finite lattice depths, which consists of a
small admixture of particle/hole pairs on top of a perfect MI. A
qualitative model based on a lowest-order calculation of the
ground state wavefunction was also presented in our previous work,
which reproduced the main trend and order of
magnitude of the observed visibility.

In the present paper, we would like to present a more precise
calculation that includes higher order corrections (see also
\cite{yu2005a}). We describe a MI state at zero temperature using
the Random Phase Approximation (RPA), already
introduced in Refs.
\cite{vanoosten2001a,dickerscheid2003a,gangardt2004a,sengupta2005a,vanoosten2005a}.
Instead of the path integral approach used by these authors, we
obtain here the RPA Green's function using a different method
inspired by Hubbard's original treatment of the fermionic model
\cite{hubbard1963,hubbard1964}. Taking the experimental geometry
and the inhomogeneous particle distribution into account, we find
good agreement with our experimental data, which provides further
support for the physical picture presented above.

The paper is organized as follows. In section \ref{BH_section}, we
recall the description of ultracold atoms in an optical lattice by
the Bose-Hubbard model, and discuss the inhomogeneous shell
structure that develops in an external confining potential.
Section \ref{IP_section} presents the calculation of the
interference pattern observed after free expansion of the atom
cloud and its link with the quasi-momentum distribution. The main
results are presented in sections \ref{maincalc} and
\ref{comparison_section}, where we respectively present the
calculation of the interference pattern in the uniform case using
the RPA, and extend it to the inhomogeneous case to compare to the
experimental data of \cite{gerbier2005a}. Details of the
calculation are described in the appendix.

\section{Bose-Hubbard hamiltonian}\label{BH_section}

In this section, we briefly recall the theoretical description of
an ultracold atomic gas trapped in an optical lattice. The optical
lattice potential, which results from the superposition of three
orthogonal and independent pairs of counterpropagating laser
beams, can be written as
\begin{equation}
V_{\rm OL}({\bf r})=V_0\left( \sin^2(k_{\rm L}x)+\sin^2(k_{\rm
L}y)+\sin^2(k_{\rm L}z)\right),
\end{equation}
Here $V_0$ is the lattice depth, $k_{\rm L}=2\pi/\lambda_{\rm L}$
is the laser wavevector, $\lambda_{\rm L}$ is the laser wavelength
and $m$ is the atomic mass. As usual, we measure $V_0$ in units of
the single-photon recoil energy $E_{\rm R}=h^2/2m\lambda_{\rm
L}^2$. The lattice potential has a simple cubic periodicity in
three dimensions, with a lattice spacing $d=\lambda_{\rm
L}/2\approx425~$nm in our case. As shown in \cite{jaksch1998a},
the behavior of the atomic system in such a potential can be
described by the Bose-Hubbard model, defined by the hamiltonian

\begin{eqnarray}\label{bosehubbard}
\mathcal{H}& = & - t \sum_{\langle i,j \rangle} \hat{a}_i^\dagger
\hat{a}_j + \sum_i \frac{U}{2} \hat{n}_i \left( \hat{n}_i-1
\right).
\end{eqnarray}
Here the operator $\hat{a}_i^\dagger$ creates an atom at site $i$,
$\hat{n}_i=\hat{a}_i^\dagger\hat{a}_i$ is the on-site number
operator, and the notation $\langle i,j \rangle$ restricts the sum
to nearest neighbors only. The relative strength between the
tunneling matrix element $t$ and the on-site interaction energy
$U$ is controlled by the depth $V_0$ of the periodic potential
which confines the atoms \footnote{In subsequent calculations we
use the approximate expressions $U\approx 5.97 (a/\lambda_{\rm
L})(V_0/E_{\rm R})^{0.88}$, and $t=1.43 (V_0/E_{\rm
R})^{0.98}\exp(-2.07\sqrt{V_0/E_{\rm R}})$. We have obtained these
formula, accurate within 1 \% in the range $V_0=8-30~E_{\rm R}$,
by numerically solving for the band structure and performing a fit
to the calculated curves.}. The phase diagram of this hamiltonian
is well known: The system is in a MI state within
characteristic lobes in a $t/U$ versus chemical potential $\mu$
phase diagram, and in a superfluid state outside of these lobes \cite{fisher1999a}.

In the experiments, an additional potential $V_{\rm ext}({\bf r})$
is superimposed to the lattice potential, leading to a spatially
varying chemical potential across the cloud. This favors the
formation of a ``wedding cake'' structure of alternating MI and
superfluid shells, which reflects the phase diagram of the
Bose-Hubbard model
\cite{fisher1999a,jaksch1998a,kashurnikov2002a,batrouni2002a}. The
external potential is due to a combination of a magnetic potential
in which the condensate is initially formed and of an optical
potential due to the Gaussian shape of the lattice beams. To a
good approximation, it can be considered as a harmonic potential
with trapping frequency
\begin{equation}\label{omega}
\Omega=\sqrt{\omega_{\rm m}^2+\frac{8 V_0}{m w^2}},
\end{equation}
where $\omega_{\rm m}$ is the frequency of the magnetic trap,
assumed isotropic, and where $w$ is the waist ($1/e^2$ radius of the intensity profile) of
the lattice beams, assumed identical for all axes. For large
lattice depths, the confinement is mainly due to the optical part.

\begin{figure}[ht!]
\includegraphics[width=8cm]{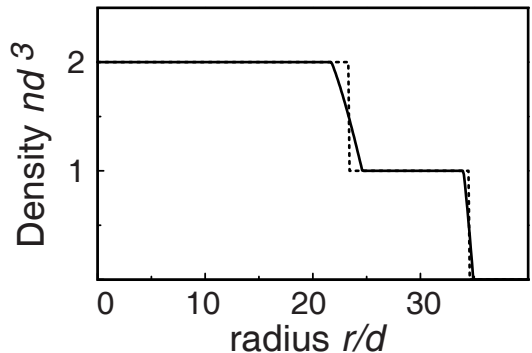}
\caption{Calculated density profile for a lattice depth
$V_0=18~E_{\rm R}$ and $N=2.2\times10^5$ atoms. Here $d$ is the
lattice spacing. The upper solid line indicates the numerical
calculation of the total density, and the dotted line is the
$U\rightarrow\infty$ extrapolation (see section
\ref{comparison_section}). } \label{density}
\end{figure}

In the current experiments, this external potential varies slowly
across the lattice. In this limit, the shell structure can be
calculated in a local density approximation, which assumes a known
relation $n_{\rm h}[\mu]$ between the density $n$ and the chemical
potential $\mu$ for the homogeneous system. Then, the
coarse-grained density \footnote{The coarse-grained density is
understood here as an average of the discrete atomic density over
a volume of linear size large compared to the lattice spacing, but
small compared to the overall extent of the cloud
\cite{kraemer2002a}.} for the inhomogeneous system is calculated
as $n({\bf r})=n_{\rm h}[\mu-V_{\rm ext}({\bf r})]$. The chemical
potential is fixed by the relation $N=\int d^{(3)}{\bf r}~n({\bf
r})$. For a fixed lattice depth and atom number, we calculate
numerically the relation $n_{\rm h}[\mu]$ using mean-field theory
at zero temperature, {\it i.e.} in the mean-field ground state
\cite{sheshadri1993a,vanoosten2001a}. We then repeat the steps
outlined above, varying the chemical potential until the target
atom number is obtained within 0.1 \%. For all calculations, the
values $\omega_{\rm m}=2\pi\times15~$Hz and $w=130~\mu$m are used,
which match our experimental parameters. We show in
Fig.~\ref{density} an example of such a calculation for a lattice
depth of $V_0=18~E_{\rm R}$ and $N=2.2\times10^5$ atoms.

The presence of the external potential significantly affects the
atom distribution in the lattice, which is determined by the
competition between interaction and potential energy. On the one
hand, expanding the cloud minimizes the density and the
interaction energy, and on the other, contracting it minimizes the
potential energy, as in conventional harmonic traps
\cite{dalfovo1999a,gerbier2004c}. The latter is favored at low
atom numbers, where only a $n_0=1$ shell forms. When more atoms
are added, the radius of the unity filled MI region increases
until a critical atom number, which we estimate to be $7\times10^4$
atoms for our parameters, is reached. Above this critical atom number, a
higher density core appears near the trap center. A MI with $2$
atoms per site is then obtained near the trap center if the
lattice depth is above the critical value $V_0\approx14.7~E_{\rm
R}$. This value has been calculated using the boundary derived in
\cite{vanoosten2001a},
\begin{equation}
\left(\frac{U}{z t} \right)_{n_0} = 2 n_0+1+2\sqrt{n_0(n_0+1)},
\end{equation}
where $z=6$ is the number of nearest neighbors in 3 dimensions. In
the specific example shown in Fig.~\ref{density}, we have chosen
$V_0=18~E_{\rm R}$ and $N=2.2\times10^5$ atoms, so that both
$n_0=1$ and $n_0=2$ MI are present. Similarly, we calculate for
our experimental parameters that an $n_0=3$ shell is also present
for atom numbers larger than $2.7\times10^5$, and lattice depths
larger than $16~E_{\rm R}$.

\section{Interference pattern}\label{IP_section}

We now turn to the description of the interference pattern
observed after release of the atom cloud from the optical lattice
and a period of free expansion. From an absorption image of such a
pattern, the phase coherence of the atomic sample can be directly
probed. The density distribution of the expanding cloud after a
time of flight $t_{\rm f}$ can be calculated as
\cite{zwerger2003a,pedri2001a,kashurnikov2002a}
\begin{equation}\label{ntof}
n({\bf r})=\left( \frac{m}{\hbar t_{\rm f}} \right)^3
\left|\tilde{w}\left(\frac{m {\bf r}}{\hbar t_{\rm f}}\right)\right|^2
\mathcal{S}\left({\bf k}=\frac{m {\bf r}}{\hbar t_{\rm f}}\right),
\end{equation}
which mirrors the momentum distribution of the original cloud.
Momentum space and real space in the image plane are related by
the scaling factor $\hbar t_{\rm f}/m$ - independent of the lattice
parameters. The envelope $\left|\tilde{w}\right|^2$ is the Fourier
transform of the Wannier function in the lowest Bloch band. When
each potential well is approximated as an harmonic potential, the
Wannier function is the corresponding gaussian ground state
wavefunction. The envelope function in Eq.~(\ref{ntof}) then reads
\begin{equation}
\left|\tilde{w}\left(\frac{m {\bf r}}{\hbar
t_{\rm f}}\right)\right|^2\approx\frac{1}{\pi^{3/2}w_{t_{\rm f}}}\exp{\left(-\frac{{\bf
r}^2}{w_{t_{\rm f}}^2}\right)},
\end{equation}
where $w_{t_{\rm f}}=\hbar t_{\rm f}/mw_0$, and where $w_0$ is the size of the
on-site Wannier function. Finally, we have defined the quantity
\begin{equation}\label{Sk}
\mathcal{S}({\bf k})= \sum_{i,j} e^{i{\bf k}\cdot({\bf r}_i-{\bf
r}_j)}\langle \hat{a}_i^\dagger \hat{a}_j\rangle.
\end{equation}
When ${\bf k}$ is restricted to the first Brillouin zone,
$\mathcal{S}({\bf k})$ is nothing else than the quasi-momentum
distribution. Information about the many-body system is contained
in this quantity, which is periodic with the periodicity of the
reciprocal lattice $2\pi/d$.  Thus, to predict the interference
pattern and compare to the experiments, our goal is to calculate
$\mathcal{S}({\bf k})$ for a given lattice depth and density.

\section{Quasi-momentum distribution in the homogeneous Mott
insulator}\label{maincalc}

For simplicity, we consider first the case of uniform filling in
the lattice, {\it i.e.} an integer number $n_0$ of atoms per site,
and we assume the system to be at zero temperature and in the
insulating phase. In the limit of zero tunneling, the ground state
wavefunction is a perfect MI, {\it i.e.} a product of number
states at each site, and its Green function $G_0$ can be
calculated exactly (see appendix). The lowest-lying excited states
of the system are ``particle'' and ``hole'' states, where a
supplementary particle is added (respectively removed) at one
lattice site. Creating these excitations costs a finite
interaction energy, respectively $E^{(+)}=U n_0$ and
$E^{(-)}=U(n_0-1)$ \cite{fisher1999a}.

To calculate the quasi-momentum distribution for a finite
tunneling $t$, many-body techniques can be applied to obtain the
single-particle Green function, $G({\bf k},\omega)$. Using a path
integral approach, several authors
\cite{vanoosten2001a,dickerscheid2003a,gangardt2004a,sengupta2005a,vanoosten2005a}
have been able to calculate the Green function of the Mott
insulator within the RPA,
\begin{equation}\label{green}
\frac{1}{\hbar}G({\bf k},\omega)=\frac{Z_{\bf k}}{\hbar \omega
+\mu -E_{\bf k}^{(+)}}+\frac{1-Z_{\bf k}}{\hbar \omega +
\mu-E_{\bf k}^{(-)}}.
\end{equation}
The poles $E_{\bf k}^{(\pm)}$ of the Green function are the
quasi-particle energies \cite{vanoosten2001a}
\begin{equation}\label{Ekpm}
E_{\bf k}^{(\pm)}=\frac{t_{\bf k}}{2}+U(n_0-\frac{1}{2}) \pm
\frac{1}{2} D_{\bf k}[n_0],
\end{equation}
In Eq.~(\ref{Ekpm}), $t_{\bf k}=-2 t \sum_{\nu=x,y,z}\cos(k_\nu
d)$ is the dispersion relation for a free particle in the
tight-binding limit, and $D_{\bf k}[n_0]=\sqrt{t_{\bf k}^2+4
t_{\bf k} U(n_0+\frac{1}{2})+U^2}$. The particle weight is $Z_{\bf
k}=\left(E_{{\bf k}}^{(+)}+U\right)/D_{\bf k}[n_0]$. In the
Appendix, we present an alternative derivation of (\ref{green})
based on the equation of motion method, which follows closely
Hubbard's method \cite{hubbard1963,hubbard1964}. Here we will
simply comment on the physical picture behind this approach. The
RPA considers that the particle/hole nature of the low-lying
excitations is not significantly changed by introducing a finite
tunneling (in technical terms, the self energy remains
approximately the same as in the $t\rightarrow 0$ limit). The
first effect of tunneling is to introduce a finite amount of
particle/hole components in the the many-body ground state
wavefunction. In the form given in \cite{gerbier2005a},
corresponding to a first order calculation, a particle/hole pair
necessarily occupies two neighboring lattice sites due to the
particular form of the tunneling hamiltonian. Through higher-order
tunneling processes captured by the Green function (\ref{green}),
the particle and the hole forming the pair can tunnel
independently. As a result, the pair acquires a mobility through
the lattice, and may even ``stretch'' over a few lattice sites.
This mobility acquired by particle/hole pairs is reflected in the
modified dispersion relation (\ref{Ekpm}), which explicitly
includes the band structure. Note finally that higher order
excitations, corresponding to occupation numbers
$n_0\pm2,n_0\pm3$, ... are neglected. At zero temperature, such
excitations become important only very close to the superfluid
transition where the MI is destroyed.

The quasi-momentum distribution can be directly deduced using the
general relation $\mathcal{S}({\bf k})=-i \lim_{\delta
t\rightarrow0^+} \int \frac{d\omega}{2\pi}~G({\bf k},\omega)e^{-i
\omega \delta t}$. Using (\ref{green}), one has
\cite{sengupta2005a}
\begin{equation}\label{Shom}
\mathcal{S}({\bf k})=n_0 \left( \frac{\frac{t_{\bf
k}}{2}+U(n_0+\frac{1}{2})}{\sqrt{t_{\bf k}^2+4 t_{\bf k}
U(n_0+\frac{1}{2})+U^2}} -\frac{1}{2} \right).
\end{equation}
To first order in $t/U$, this reduces to
\begin{equation}\label{S1}
\mathcal{S}({\bf k}) \approx n_0-2n_0(n_0+1)t_{\bf k}/U,
\end{equation}
also obtained in \cite{gerbier2005a} by calculating the many-body
wave function perturbatively. We find that the two predictions
rapidly converge. For example, they differ by less than 10 \% for
$U/zt>6.6,11.6$ and $16.9$ for $n_0=1,2$ and $3$, respectively.
These values have to be compared to the respective critical values
for MI formation, $U/zt=5.83,9.89,13.93$. This indicates that the
coherence beyond nearest neighbors is rather rapidly lost as one
goes further into the MI phase. However, the visibility itself
remains finite in a substantial range of $U/zt$, implying a
persistent short-range coherence.

\section{Comparison with the experiments}\label{comparison_section}

To compare with the experiments reported in \cite{gerbier2005a},
several features have to be taken into account. First, only the
column density $n$ is accessible experimentally, {\it i.e.} the
density integrated along the probe line-of-sight (which we take
here parallel to the $z$ axis). Second, the visibility is
experimentally deduced from two points according to
\begin{equation}\label{V}
\mathcal{V}_{\rm exp}=\frac{n_{\rm max}-n_{\rm min}}{n_{\rm
max}+n_{\rm min}}.
\end{equation}

\begin{figure}[ht!]
\includegraphics[width=8cm]{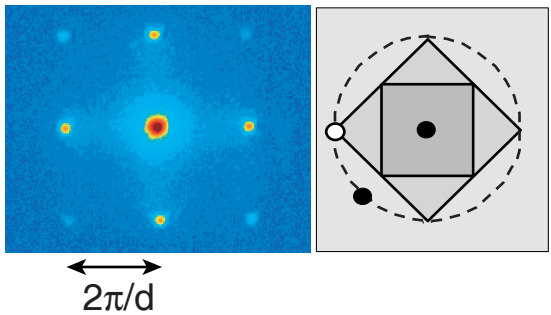}
\caption{Measurement of visibility. The interference pattern shown
in the left graph corresponds to a lattice depth of $8~E_{\rm R}$,
in the superfluid regime. The right graph indicates the geometry
of the reciprocal lattice. Gray areas are the first and second
Brillouin zones (projected in the image plane), and the white dot
indicate the position of the maxima of the interference pattern.
Along the circle, the Wannier function envelope takes the same
value, and we measure the interference ``minimum'' at the
intersection of this circle and of the diagonal of the lattice
square lattice, indicated by the black dot.} \label{expvis}
\end{figure}

To eliminate the Wannier envelope, $n_{\rm max}$ and $n_{\rm min}$
are measured from two points at the same distance from the cloud
center (see Fig.~\ref{expvis}), so that the envelope automatically
cancels out. For example, $n_{\rm max}$ is found at point
$(2\pi/d,0)$ and $n_{\rm min}$ at
$(2\pi/\sqrt{2}d,2\pi/\sqrt{2}d)$. This reduces the visibility
compared to the usual definition. In the theoretical calculation
it is straightforward to account for these two effects.

The third effect, the shell structure of the MI, is handled here
in an approximate way. In the numerical calculations, the shell
distribution always includes small regions with non-integer
filling, which the theory above cannot handle. However, these
domains are small, and have a strongly depleted superfluid
component, so that we do not expect them to have a large effect on
the visibility. Therefore, we approximate the density distribution
by a ``ziggurat''-like profile, where only MI shells are present.
The actual extension of each shell is calculated as if $t$ were
zero, taking the external potential into account
\cite{demarco2005a}. In Fig.~\ref{density}, we compare the profile
in this approximation (dotted line) with the numerically
calculated one (solid line). For large lattice depths, both agree
reasonably. Note that the density profile still depends weakly on
the lattice depth through the external confinement [see Eq.
(\ref{omega})].

\begin{figure}[ht!]
\includegraphics[width=8cm]{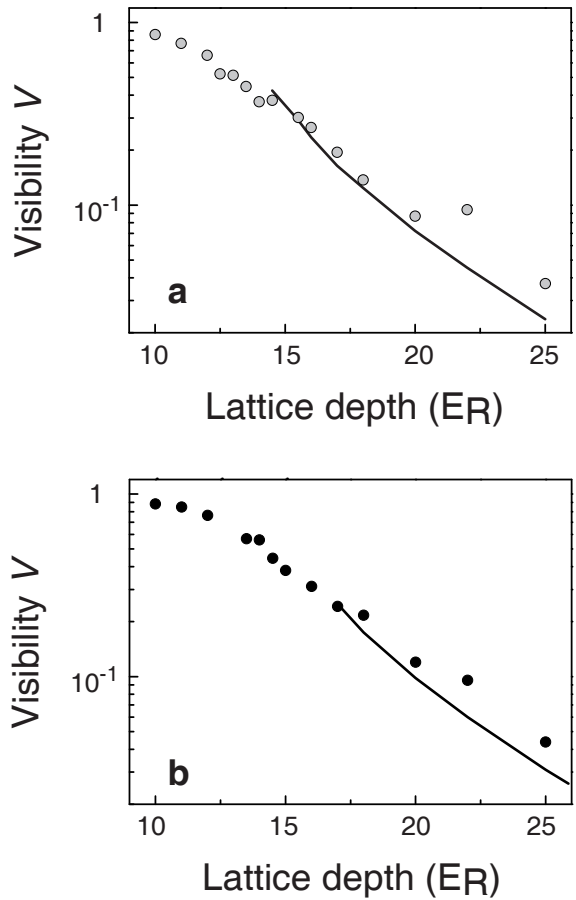}
\caption{Comparison between the measured and the calculated
visibility. The upper and lower graphs correspond to atom numbers
$N=2.2\times10^5$ and $N=5.6\times10^5$, respectively. The dotted
and dashed lines indicate the calculated visibility for
homogeneous MI with filling factor $n_0=1,2$. The solid lines are
calculations including the inhomogeneous shell distribution.
Typical standard deviations for the experimental data are 1 \% or
below. Our calculation of the equilibrium distribution at zero
temperature indicate that in case {\bf a}, only MI regions with
$n_0=1$ and $n_0=2$ atoms per site form, whereas in case {\bf b},
a core with $n_0=3$ is also present.} \label{comparison}
\end{figure}

The momentum distribution deduced from Eqs.
(\ref{ntof},\ref{Shom}) is averaged over the distribution of atoms
to compare with the experimental data (see \cite{gerbier2005a} for
details on the experiment). The results are plotted versus lattice
depth in Fig.~\ref{comparison}, for two different atom numbers in
the lattice. For the lowest atom number $N=2.2\times10^5$, we
calculate that only $n_0=1$ and $n_0=2$ shells are present. For
the largest $N=5.6\times10^5$, a core with $n_0=3$ atoms per site
is also present. Note that in the latter case, the actual density
distribution might deviate more from the calculated one, due to
three-body losses in the $n_0=3$ region. We find that the
calculation agrees with the measured visibility within 20 \% for
$V_0\leq22~E_{\rm R}$. The theory curves terminate when the MI
shell with highest filling disappears, as it is replaced by a
large superfluid core not described by our theory. Note that the
calculation does not include any free parameter.

\begin{figure}[ht!]
\includegraphics[width=8cm]{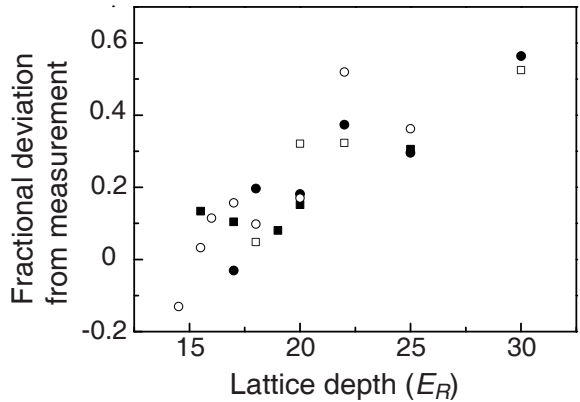}
\caption{Fractional deviation of the calculated visibility from
the measurements. Symbols denote different data sets with
different atom numbers and densities ($\square$: $N=2.2 \times
10^5$, $\blacksquare$: $N=3.6 \times 10^5$, $\medcirc$: $N=4.3
\times 10^5$, $\medbullet$: $N=5.9 \times 10^5$).}
\label{deviation}
\end{figure}

However, we consistently find that the calculated value lies below
the measured visibility for large lattice depths $V_0\leq22~E_{\rm R}$.
Moreover, the deviation increases with increasing lattice depths, which
shows that the superfluid shells play little role in determining the visibility for such large
lattice depths, as assumed in our calculation. In
Fig. \ref{deviation}, the fractional deviation of the
calculated visibility from the measured one is plotted versus
lattice depth for four data sets. Remarkably, although the atom
numbers are rather different from one data set to another we find
a common trend in the data. On the other hand,
this observation also suggests that a breakdown of adiabaticity
occurs for the particular ramp used in the experiments to increase
the laser intensity to its final value, a point already identified
in \cite{gerbier2005a}. We conclude that, perhaps surprisingly,
the visibility in the MI may be a sensitive probe of the many-body
dynamics of the superfluid-to-insulator transition.

\section{Conclusion}

In conclusion, we have derived in this paper a theoretical
expression for the interference pattern of a Mott insulator after
release from the optical lattice and a time of flight. Our
calculations take deviations from perfect filling due to a finite
tunneling into account, and use a simplified but realistic model
of the shell structure of the MI. Good agreement with our
experimental data reported in \cite{gerbier2005a} is found, at
least for moderate lattice depths. For very large lattice depths,
an increasing deviation points to non-adiabatic effects in the
conversion from a condensate to an insulating state, which could
in principle be studied by the method presented here.
Nevertheless, in view that no free parameter is included in the
theory, we conclude that the momentum distribution (\ref{Shom})
describes the system well. This supports the physical picture of
the system as a (dilute) gas of partice/hole pairs, mobile through
the lattice, on top of a regularly filled Mott insulator.
Furthermore, the validity of the RPA to describe their behavior is
qualitatively verified.

Our calculation neglects entirely the superfluid component, which
is correct only for large lattice depths where the system is
almost completely insulating. Recently, several authors
\cite{garciaripoll2004a,schroll2004a,sengupta2005a} have proposed
to modify the standard mean-field description
\cite{fisher1999a,sheshadri1993a} to better account for long- and
short-range coherence. It would be interesting to compare the
predictions of those approaches with our data for lower lattice
depths, where the system is expected to be a strongly depleted
superfluid, and therefore amenable neither to a Bogoliubov-like
description nor to a strongly interacting one as provided in this
paper. Also, an investigation of finite temperature effects
\cite{dickerscheid2003a} would be useful. In particular, an
interesting question is to know whether the visibility
measurements presented here could be used for thermometry in the
lattice.

We would like to thank Dries van Oosten, Paolo Pedri and Luis
Santos for useful discussions. Our work is supported by the
Deutsche Forschungsgemeinschaft (SPP1116), the European Union
under a Marie-Curie Excellence Grant and AFOSR. FG acknowledges
support from a Marie-Curie Fellowship of the European Union.

\appendix
\section{Green function of the homogeneous Mott insulator in the
Random Phase approximation}\label{calc_appendix}

In this Appendix, we present a derivation of Eq.~(\ref{green})
using the equation of motion approach. The single-particle Green
function is defined at zero temperature as
\begin{eqnarray}\label{Gdef}
G_{ij}(t) & = & -i \langle \mathcal{T}\hat{a}_{i}(0) \hat{a}_{
j}^\dagger(t) \rangle\\\nonumber &=& -i \theta(t) \langle
\hat{a}_{i}(0) \hat{a}_{ j}^\dagger(t) \rangle\\ \nonumber & & -i
\theta(-t) \langle \hat{a}_{ j}^\dagger(t) \hat{a}_{i}(0)\rangle,
\end{eqnarray}
where $\mathcal{T}$ is the time-ordering operator and $\theta$ is
the Heaviside step function. Since we consider a time-independent
and homogeneous system, we take a Fourier transformation of this
equation with respect to space and time (denoted by the symbol
$\mathcal{F}$), and define
\begin{eqnarray}
G({\bf k},\omega) & = & \mathcal{F}\left[G_{ij}(t)\right]
\end{eqnarray}

In the frequency-momentum representation, the Heisenberg equation
of motion $i \hbar \frac{\partial}{\partial
t}G_{ij}(t)=\left[H,G_{ij}(t)\right]$ takes the form
\begin{eqnarray}\nonumber
(\hbar \omega+\mu) G({\bf k},\omega) & = & 1 + t_{\bf k}G({\bf
k},\omega)\\ \label{Gk} & & - i \mathcal{F}\left[\langle
\mathcal{T} n_i\hat{a}_{i}(0) \hat{a}_{ j}^\dagger(t) \rangle
\right].
\end{eqnarray}
The last term on the right hand side of (\ref{Gk}) is usually
rewritten as $\Sigma({\bf k},\omega)G({\bf k},\omega)$, where
$\Sigma$ is the self-energy. This gives the expression
\begin{eqnarray}
G({\bf k},\omega) & = & \frac{1}{\hbar\omega+\mu -t_{\bf
k}-\Sigma({\bf k},\omega)}.
\end{eqnarray}

Let us first assume that no tunneling is present ($t=0$). In this
case, the Green function can be obtained exactly from its
definition (\ref{Gdef}) and the ground state wave function
$|\Psi\rangle = \prod_i |n_0\rangle_i$, where each site is in the
Fock state $|n_0\rangle$. The result $G_0(\omega)$ is independent
of momentum, and reads
\begin{equation}\label{G0}
G_0(\omega)=\frac{n_0+1}{\hbar\omega+\mu-U
n_0}-\frac{n_0}{\hbar\omega+\mu-U (n_0-1)}.
\end{equation}
In this self-interacting limit, we can rewrite Eq. (\ref{G0}) as
$G_0(\omega)^{-1}=\hbar\omega+\mu-\Sigma_0(\omega)$, with the self
energy \cite{vanoosten2005a}
\begin{equation}
\Sigma_0(\omega)=2Un_0-\frac{U^2 n_0(n_0+1)}{\hbar \omega+\mu+U}.
\end{equation}
This expression is exact in the $t\rightarrow0$ limit, and
coincides with the one found in \cite{vanoosten2005a}. The first
term is simply the Hartree-Fock energy per particle for
uncondensed atoms (hence the factor of $2$), whereas the second
term - which has the same order of magnitude at low energy -
accounts for the correlations between particle that drive the
system into the perfectly ordered ground state.

If we now restore a finite tunneling, but still consider a system
in the insulating phase, a reasonable approximation is to assume
that the self energy is not changed with respect to the strongly
interacting limit. We comment on this approximation in the test.
Making this approximation yields
\begin{eqnarray}\nonumber
G({\bf k},\omega) & \approx & \frac{1}{\hbar\omega+\mu -t_{\bf
k}-\Sigma_0(\omega)}\\ \label{green2} & = &
\frac{G_0(\omega)}{1-t({\bf k})G_0(\omega)},
\end{eqnarray}
which has a typical RPA form. Using the explicit result for $G_0$,
 we obtain after some algebra Eq. (\ref{green}) in the text,
 which explicitly displays particle and hole components.

%
%

\end{document}